\documentclass[epj]{svjour}
\usepackage{graphics}
\usepackage{graphicx}
\usepackage{epsfig}

\newcommand{\beq}{\begin{eqnarray}}
\newcommand{\eeq}{\end{eqnarray}}
\newcommand{\be}{\begin{eqnarray*}}
\newcommand{\ee}{\end{eqnarray*}}

\newcommand{\D}{{\cal D}}
\newcommand{\Pom}{{\hspace{ -0.1em}I\hspace{-0.25em}P}}
\def\lsim{\raise0.3ex\hbox{$<$\kern-0.75em\raise-1.1ex\hbox{$\sim$}}}
\def\gsim{\raise0.3ex\hbox{$>$\kern-0.75em\raise-1.1ex\hbox{$\sim$}}}
\begin{document}
\title{Gluon shadowing in the Glauber-Gribov model}
\author{K. Tywoniuk\inst{1} \and I.~C.~Arsene\inst{1} \and
  L. Bravina\inst{1,2} \and A.~B.~Kaidalov\inst{3} \and E.~Zabrodin\inst{1,2}
}                     
\institute{Department of Physics, University of Oslo\\
  0316 Oslo, Norway
  \and
  Institute of Nuclear Physics, Moscow State University\\
  RU-119899 Moscow, Russia
  \and
  Institute of Theoretical and Experimental Physics\\
  RU-117259 Moscow, Russia
} \mail{konrad.tywoniuk@fys.uio.no}
\date{Received: date / Revised version: date}
%
\abstract{
  New data from HERA experiment on (diffractive) deep inelastic scattering has been
  used to parameterize nucleon and Pomeron structure functions. Within
  the Gribov theory, the parameterizations were employed to
  calculate gluon shadowing for
  various heavy ions and compared our results with predictions from
  other models. Calculations for d+Au
  collisions at forward rapidities at ultra-relativistic energies have
  been made and are compared to RHIC data on the nuclear modification
  factor. Results for gluon shadowing are also confronted with recent
  data on the nuclear modification factor at $\sqrt{s} = 17.3$ GeV at
  various values of the Feynman variable $x_F$, and the energy
  dependence of the effect is discussed.
  \PACS{
    {12.40.Nn}{Regge theory, duality, absorptive/optical models}   \and
    {13.60.Hb}{Total and inclusive cross sections (including deep
      inelastic processes)}   \and
    {13.85.-t}{Hadron-induced high- and super-high-energy
      interactions (energy $>$ 10 GeV)}       \and
    {25.75.-q}{Relativistic heavy-ion collisions}
  } 
} 
\maketitle
\section{Introduction}\label{intro}

The fact that the nuclear structure function, $F_2^{A}$, per number
of constituent nucleons is smaller than the structure function of a
single nucleon, $F_2^{N}$, for $x<0.1$ \footnote{In the infinite
momentum frame, $x$ is the longitudinal momentum fraction of a
parton in the nucleon} is one of the most intriguing effects in
modern high-energy nuclear physics. This effect is called nuclear
shadowing. We define the nuclear ratio \beq \label{eq:Rdefinition}
R\left( \frac{A}{N} \right) \;=\; \frac{F_2^{A} (x, Q^2)}{A \,
  F_2^{N} (x, Q^2)}
\eeq which is smaller than unity in the shadowing region. The
nuclear ratio has been measured for many nuclei and reveals also an
interesting structure for $x > 0.1$ (see \cite{arneodo} for further
references). In high-energy hadron-nucleus collisions we probe the
partonic structure of the nucleus at low values of $x$. For a given
energy the smallest available values of $x$ of the partons in the
nucleus are to be found in the fragmentation region, where $x_F =
p_z / p^{max}_z$ of the observed particle is close to 1, which
corresponds to large pseudorapidity $\eta$.

An understanding of the so-called cold nuclear effects, or initial
state effects, in hadron-nucleus collisions are therefore an
important benchmark for nucleus-nucleus collisions.

A significant change in the underlying dynamics of a hadron-nucleus
collision takes place with growing energy of the incoming particles.
At low energies, the total cross section is well described within
the probabilistic Glauber model \cite{glauber}, which only takes
into account elastic rescatterings of the initial hadron on the
various nucleons of the nucleus. Elastic scattering is described by
Pomeron exchange. At higher energies, $E > E_{crit} \sim
m_{\scriptstyle{N}} \mu R_A$ corresponding to a coherence length
\beq \label{eq:cohlenght} l_C \;=\; \frac{1}{2 \, m_N \, x} \;, \eeq
the typical hadronic fluctuation length can become of the order of,
or even bigger than, the nuclear radius, $R_A$, and there will be
coherent interaction of constituents of the hadron with several
nucleons of the nucleus. The sum of all diagrams was calculated by
Gribov \cite{gribov1}. In this framework, the diffractive
intermediate states has to be accounted for in the sum over
subsequent rescatterings. The space-time picture analogy to the
Glauber series is lost, as the interactions with different nucleons
of the nucleus happens instantaneously. The phenomenon of coherent
multiple scattering is referred to as shadowing corrections.

An additional effect which comes into play at high energies, is the
possibility of interactions between soft partons of the different
nucleons in the nucleus. In the Glauber-Gribov model this
corresponds to multi-Pomeron interactions. These diagrams are called
enhanced diagrams \cite{kaidalov_qm}, and can also be understood as
interactions between strings formed in the collision. E.g. the
triple-Pomeron vertex is proportional to $A^{1/3}$ in hA collisions,
and so it becomes very important for collisions on very heavy
nuclei.

In what follows we will describe a model for hadron-nucleus
collisions in Section \ref{sec:model}. In Section \ref{sec:exp} we
will discuss the experimental data on both inclusive and diffractive
deep inelastic scattering (DIS) cross-sections measured at HERA used
as input to the model. In Section \ref{sec:Rnumer} we will present
the results for the shadowing ratio, and also compare them to recent
experiments at the Relativistic Heavy Ion Collider (RHIC) in Section
\ref{sec:rhic}. In Section \ref{sec:sps} we compare our calculations
to recent measurements done at much lower energies, $\sqrt{s} =
17.3$ GeV, and discuss the energy dependence of the shadowing
effect. We summarize and conclude in Section \ref{sec:concl}.

\section{The Model} \label{sec:model}
\begin{figure}[b]
  \centering
  \includegraphics[width=0.8\linewidth]{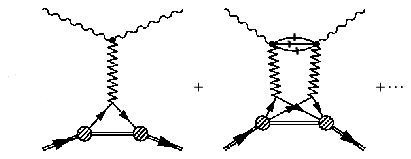}
  \caption{\label{fig:diagram} The single and double scattering from
  to the total cross section (the diffractive intermediate states are
  on mass shell).}
\end{figure}
We will assume that the nucleus consists of $A$ independent nucleons,
in the spirit of the Glauber model.
The scattering amplitude of an incoming hadron on a nuclear target,
can be expanded in a multiple scattering series \cite{cap98}
\beq
\label{eq:sum}
\sigma_A \;=\; A\sigma_N \,+\, \sigma_A^{(2)} \,+\, ...\;,
\eeq
where the first term is simply the Glauber elastic rescattering (see
Fig.~\ref{fig:diagram}).
The second term in (\ref{eq:sum}) is related to
diffractive DIS through the AGK
cutting rules \cite{agk}. It is given by \cite{gribov2}
\beq
\label{eq:second}
\nonumber
\sigma^{(2)}_A \;&=\;& -4\pi \, A(A-1) \, \int \mbox{d}^2b \, T^2_A(b)
\\ && \times\;\int_{M^2_{min}}^{M^2_{max}} \, \mbox{d}M^2
\left[\frac{\mbox{d}\sigma^\D_{hN}}{\mbox{d}M^2
    \mbox{d}t}\right]_{t=0} F^2_A(t_{min}) \;,
\eeq where $M^2$ is the mass of the diffractively produced
intermediate state, $T_A (b)$ is the normalized nuclear thickness
function and $\mbox{d}\sigma^\D_{hN} / \mbox{d}M^2 \mbox{d}t$ is the
differential cross section for diffractive dissociation of the
hadron. In the second integral, $M^2_{max} = Q^2(x_\Pom^{max}/x -1)$
is found by demanding a large rapidity gap in the diffractive
dissociation. Calculations are made both for $x_\Pom^{max} = 0.1$ as
in \cite{cap98} and for $x_\Pom^{max} = 0.03$ as in \cite{fgs03},
although the former is more convenient as it guarantees the
disappearance of nuclear shadowing at $x \sim 0.1$ as in
experimental data. Coherence effects are taken into account in the
form factor \beq \label{eq:formfact} F_A (t_{min}) \;=\; \int
\mbox{d}^2b \, J_0(b\sqrt{-t_{min}})\, T_A(b) \eeq which is equal to
1 at $x \rightarrow 0$ and decreases with increasing $x$ due to the
loss of coherence for $x \geq (2 m_N R_A)^{-1}$, see
(\ref{eq:cohlenght}) (here we have put $t_{min} = -m_N^2 x_\Pom^2$).

The second order elastic cross section in (\ref{eq:second}) is
obviously negative and will lead to a reduction of the total cross
section. In the small $x$ region, it is also necessary to include
higher orders terms in (\ref{eq:sum}) in order not to violate
unitarity of the total cross section, as was noted in
\cite{kaidplb}.

Summation of all terms in (\ref{eq:sum}) is model dependent. The
Schwimmer unitarization \cite{schwimmer} for the total $h A$ cross
section, which also sums up all Pomeron tree diagrams, is used to
obtain \beq \label{eq:sch} \frac{\sigma_{\gamma*
\scriptscriptstyle{A}}^{Sch}}{A \,\sigma_{\gamma^*
    \scriptscriptstyle{N}}}  \;=\; \int
\mbox{d}^2b \; \frac{T_A (b)}{1 \,+\, (A-1) f(x, Q^2) T_A (b) } \;,
\eeq where $f(x,Q^2)$ is the effective shadowing function. Following
\cite{cap01,H1abs} in choice of parameters and factorization, one
can get the shadowing function as \beq \label{eq:fsimpl} f(x, Q^2)
\;=\;& 4\pi\; \int_x^{x_\Pom^{\scriptscriptstyle{max}}}
\mbox{d}x_\Pom &B(x_\Pom)\, \frac{F_{2 \D}^{(3)}(x_\Pom, Q^2,
\beta)}{F_2 (x, Q^2)}
\\ \nonumber
 & & \times F_A^2(t_{min}) \;.
\eeq
Here $B(x_\Pom) = 0.184 - 0.02 \ln \left(x_\Pom
\right)\,\mbox{fm}^2$. The lhs. in (\ref{eq:sch}) is defined as the
shadowing ratio $R^{Sch}\left(A/N \right) (x)$.

The structure function $F_2$ and the diffractive structure function
$F_{2\D}$ of the single nucleon are taken as input from experiment.
The extension to the nuclear case is therefore parameter-free except
for the unitarity constraints leading to eq. (\ref{eq:sch}). This is
a remarkable feature of the Glauber-Gribov model \cite{armestorev}.

The model is valid for low values of $x \leq 0.01$ and intermediate
$Q^2 < 10 \mbox{ GeV}^2$. In what follows, we will neglect the $Q^2$
dependence of the model, knowing that the presence of a strong
$Q^2$-dependent term is not required to describe nuclear data at low
$Q^2$ which is relevant for our present considerations \cite{cap98}.

\section{Inclusive and diffractive data}
\label{sec:exp}

In deep inelastic scattering (DIS) the structure function of a
nucleon is related to the total cross section of $\gamma^* N$
interaction through factorization at high scales valid in
perturbative QCD. The structure function holds information about the
partonic content of the nucleon, and is given by a sum of parton
distribution functions (PDFs) \beq \label{eq:structfuncdef} F_2
\left( x, Q^2 \right) \;\propto\; \sum_{i=g,u,d,s...} \, x f_i
\left( x, Q^2 \right) \;, \eeq where the sum is over all types of
partons. Similar to the inclusive DIS case, a factorization theorem
has been proved in perturbative QCD to hold for diffractive
structure functions \cite{collins}. The diffractive structure
function $F_{2\D}^{(3)}$ in eq. (\ref{eq:fsimpl}) is given by \beq
\label{eq:pomstructfunc} F_{2\D}^{(3)} \, \left( x_\Pom, Q^2, \beta
\right) \;=\; \overline{f}_\Pom \left( x_\Pom \right) \, F_2^\Pom
\left( \beta, Q^2 \right) \;, \eeq where $\overline{f}_\Pom$ is the
$t$-integrated Pomeron flux and we have assumed so-called Regge
factorization. The Pomeron flux factor is defined as \beq
\label{eq:pomeronflux} \overline{f}_\Pom \left(x_\Pom \right) \;=\;
\int^{t_{min}}_{t_{cut}} \, \frac{e^{B_0 \, t}}{x_\Pom^{2
\alpha_\Pom (t) - 1}} \, \mbox{d}t \;, \eeq where we assume a linear
Pomeron trajectory, $\alpha_\Pom (t) =$ $ \alpha_\Pom (0) +
\alpha'_\Pom t$. The values of the parameters are taken to be
$\alpha_\Pom (0) = 1.173$ and $\alpha'_\Pom = 0.26 \mbox{
  GeV}^{-2}$, and we put $B_0 = 4.6 \mbox{ GeV}^{-2}$ (see
\cite{baronepredazzi} for details). $F_2^\Pom$ is called the
structure function of the Pomeron, and so it is possible to
introduce a partonic structure of the Pomeron
\cite{ingelmanschlein}, as in eq. (\ref{eq:structfuncdef}).

For both inclusive and diffractive DIS, the gluon content clearly
dominates over the quark one at low $x$ and intermediate $Q^2$
values relevant to our present considerations. In what follows we will
only consider the gluon parton
distribution functions of the nucleon and Pomeron. Quark shadowing was discussed
in \cite{cap98}. The ratio under the integral in eq. (\ref{eq:fsimpl})
can, in other words be understood as the density of gluons in the
Pomeron compared to the density of gluons in the nucleon!

For the nucleon structure function we will use the next to leading
order (NLO)
ZEUS-S QCD fit of the gluon PDF \cite{chekanov03} at
$Q^2 = 7 \mbox{ GeV}^2$.
\begin{figure}[!b]
  \centering
  \includegraphics[width=0.8\linewidth]{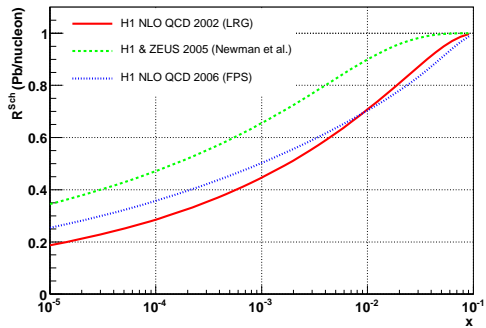}
  \caption{\label{fig:paramcomp} Nuclear shadowing ratio $R
  (\mbox{Pb}/\mbox{N})$ for different parameterizations of the
  diffractive gluon parton distribution function.}
\end{figure}
The gluon diffractive parton distribution function (dPDF) was
measured for intermediate $Q^2$ at both HERA experiments: ZEUS and
H1. The situation regarding the gluon content of the Pomeron is at
present rather uncertain with a large discrepancy between the
presented results. The origin of the discrepancy is unknown. We will
make calculations for three different fits which we will briefly
mention
\begin{itemize}
\item H1 NLO QCD 2002 fit \cite{H1abs} (large-rapidity-gap data,
  gives also a good description of heavy quark and dijet production)
\item H1 \& ZEUS 2005 fit \cite{H1ZEUS} (H1 NLO QCD analysis of ZEUS
  $M_X$-data)
\item H1 NLO QCD 2006 fit \cite{H12006_1,H12006_2} (forward
  proton spectrometer data, fit A was used)
\end{itemize}
The gluon dPDF for the Pomeron was parameterized at fixed $Q^2 = 6.5
\mbox{ GeV}^2$ ($Q^2 = 8.5 \mbox{ GeV}^2$ in the second case).
Both inclusive and diffractive distributions were fitted by a simple
function
\beq
x \, f_g \left( x, Q^2 \right) \;=\;  x \, f_g \left( x \right) \;=\;
A x^{-\delta} \left(1-x \right)^\gamma \;,
\label{eq:fitfunc}
\eeq
where $A$, $\delta$ and $\gamma$ are fitting parameters.

\section{Nuclear shadowing ratio}\label{sec:Rnumer}

Numerical calculations of the nuclear shadowing ratio for Pb, defined in
eq. (\ref{eq:sch}), for different parameterizations of the gluon dPDF
as described in the previous section are presented in
Fig.~\ref{fig:paramcomp}.
The fit to ZEUS data \cite{H1ZEUS} predicts weakest gluon shadowing,
while the fits to the both of the H1 datasets are compatible with each
other and predict a much stronger shadowing effect. The strongest
gluon shadowing is obtained for the H1 NLO QCD 2002 fit \cite{H1abs},
which is almost twice as big as the ZEUS one \cite{H1ZEUS} for the
whole range of x.

\begin{figure}[t]
  \begin{minipage}[t]{\linewidth}
    \centering
    \includegraphics[width=0.8\linewidth]{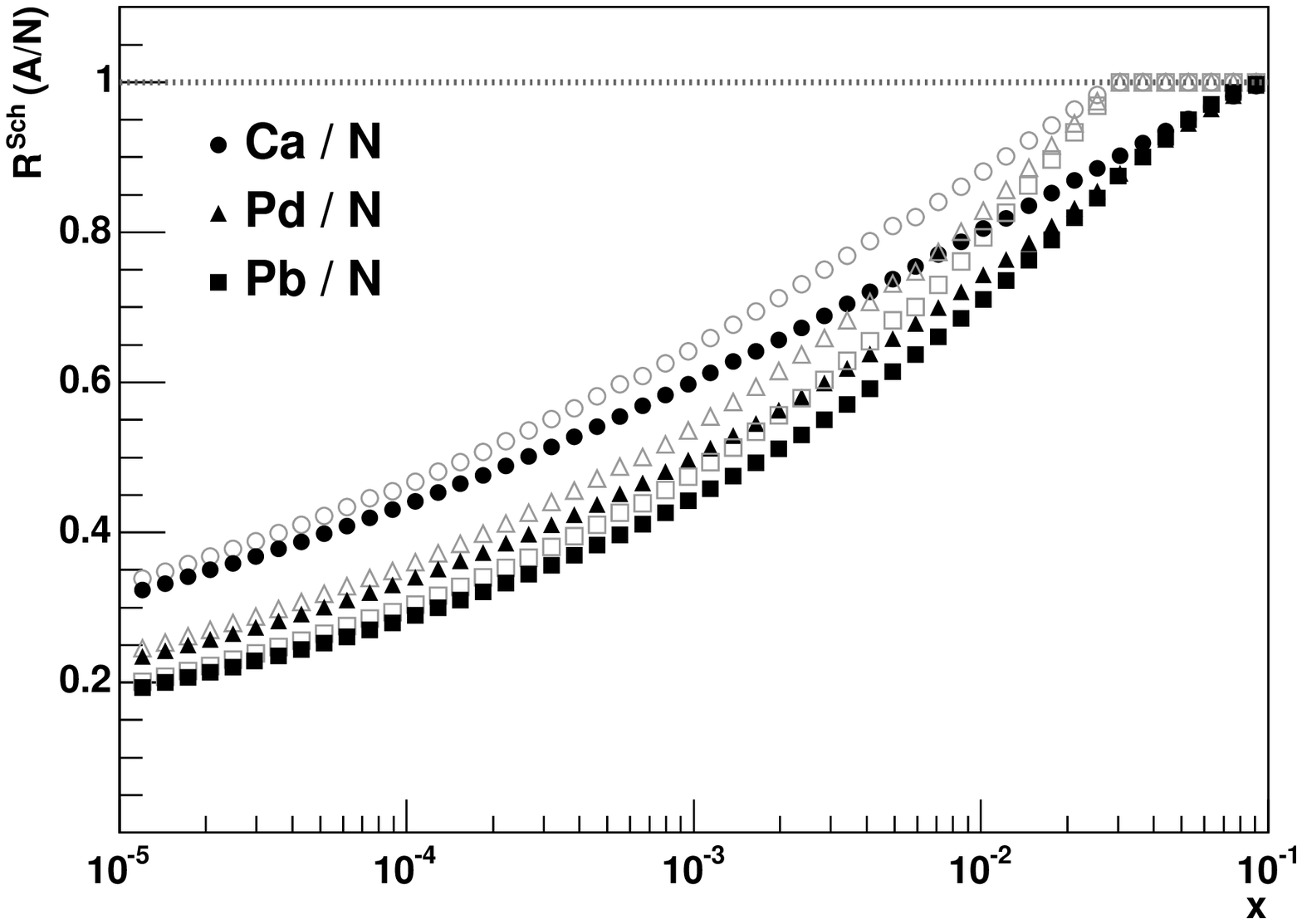}
    \caption{Gluon shadowing for heavy ions. Closed (open) symbols are for
      $x_\Pom^{max} = 0.1$ ($0.03$).}
    \label{fig:AN}
  \end{minipage}
  \begin{minipage}[t]{\linewidth}
    \centering
    \includegraphics[width=0.8\linewidth,height=1.81in]{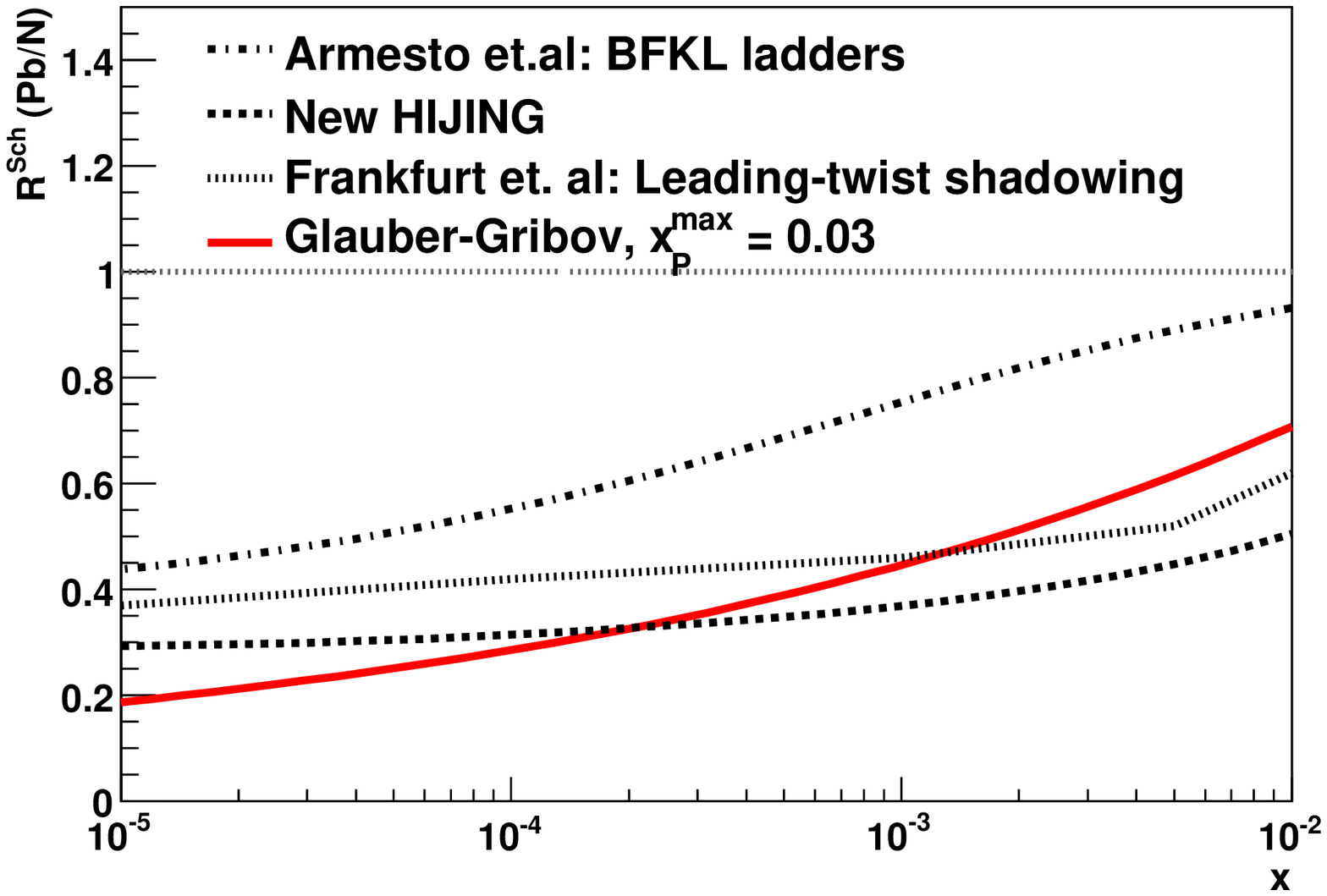}
    \caption{Comparison of our results the Pb/nucleon ratio with other models, at fixed $Q^2$.}
    \label{fig:comp}
  \end{minipage}
\end{figure}

Gluon shadowing for various heavy ions (Ca, Pd and Pb) calculated
with (\ref{eq:sch}) is presented in Fig.~\ref{fig:AN} (H1 NLO QCD
2002 fit \cite{H1abs} is used). The effect is strong at small $x$,
and disappearing at $x = x_\Pom^{max}$. This is a consequence of the
coherence length in the form factor (\ref{eq:formfact}), and the
vanishing integration domain in (\ref{eq:fsimpl}). Gluon shadowing
is as low as 0.2 for the Pb/N ratio at $x \sim 10^{-5}$.

A comparison of our results for Pb/nucleon ratio at $Q^2 = 6.5
\mbox{ GeV}^2$ with $x_\Pom^{max} = 0.03$, with other models,
calculated at $Q^2 = 5 \mbox{ GeV}^2$, is presented in
Fig.~\ref{fig:comp} (H1 NLO QCD 2002 fit \cite{H1abs} is used). In
\cite{armesto02}, the authors have calculated shadowing within the
BFKL formalism \cite{bfkl}, while \cite{hijing} is based on a
parameterization of pp-data. The authors of \cite{fgs03} make their
calculations within a similar framework as the presented model.

For $x \leq 10^{-3}$ our model predicts stronger gluon shadowing
compared to \cite{armesto02} (dashed-dotted line) and \cite{fgs03}
(dotted line), while \cite{hijing} (dashed line) predicts the
strongest effect down to $x \sim 10^{-4}$. Our calculations are
close to the predictions of \cite{fgs03} for $x > 10^{-3}$ for this
choice of $x_\Pom^{max}$, while we are consequently below the
predictions of \cite{armesto02}.

\section{Shadowing effects in d+Au collisions}
\label{sec:rhic} The model is now employed to study particle
production in d+Au collisions at RHIC energy $\sqrt{s} = 200$ GeV.
There has been observed an increasing suppression of the nuclear
modification factor (NMF) \beq \label{eq:nmf} R_{d Au} \;=\;
\frac{1}{\left< N_{coll} \right>} \, \frac{\mbox{d}^2N^{d Au}\, / \,
\mbox{d} p_T \mbox{d}
    \eta}{\mbox{d}^2 N^{pp}_{inel} \,/\,\mbox{d}p_T \mbox{d} \eta}
\eeq with increasing pseudorapidity of the observed particle
\cite{brahms}. Various models have been utilized to explain the
observed features, the most successful of which is the
Color-Glass-Condensate (CGC) model \cite{kharzeev2003}, which
assumes gluon saturation for the kinematical domain reached in
hadron-nucleus collisions at RHIC. It is instructive to point out,
that the model presented here does not assume a saturation scale,
like in \cite{kharzeev2003}. Yet, as we have already mentioned,
Pomeron interactions are taken into account to preserve unitarity of
the scattering amplitude. In this sense it should give similar
results as models assuming gluon ladder fusion.

In the Glauber-Gribov model, the multiplicity reduction due to
shadowing compared to the simple Glauber model is easily obtained in a
factorized form. The theoretical prediction is given by \cite{cap99}
\beq
\label{eq:shadow}
R^{theo}_{d Au} \;&=&\; R^{Sch}_d (x_{p}) R^{Sch}_{Au} (x_{t}) \;,
\eeq
where the deuteron will be treated as a point
particle in impact parameter space, but with the shadowing
found from (\ref{eq:sch}). The collision is described by the
following jet kinematics
\beq
\label{eq:jet}
 x_{p (t)} \;=\; c\,
p_T \, e^{\pm \eta}/ \sqrt{s} \;,
\eeq
for the projectile (target) $x$-value respectively.
In eq. (\ref{eq:jet}) $p_T$ is the transverse
momentum of the particle, and we assume that most of the
high-$p_T$ particles come from jets
$c$ times more energetic.

An important and well-known effect that is not taken into account in
the model presented here, is the Cronin effect
\cite{cronin1,cronin2}, or $p_T$-broadening of the produced
particles, which leads to an effective enhancement of the NMF seen
at midrapidity for $p_T > 2$ GeV \cite{brahms,phenix}. In what
follows we will assume that this effect stays the same for all
pseudorapidities.

\begin{figure}[!t]
  \centering
  \includegraphics[width=1.\linewidth]{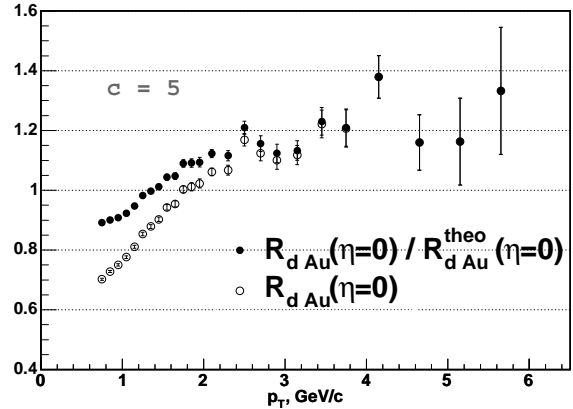}
  \caption{Shadowing effects for d+Au collisions $\sqrt{s} = 200$ GeV
  at midrapidity for $c=5$.}
  \label{fig:eta0}
\end{figure}

We therefore extract the gluon shadowing effects in the NMF at $\eta
= 0$ by defining $R^{\mathit{norm}}_{d Au} = \left[R_{d
Au}^{\mathit{exp}}
 / R_{d Au}^{\mathit{theo}}\right]_{\eta=0}$, shown in
Fig.~\ref{fig:eta0} for $c = 5$.
As expected, the shadowing dies out at high $p_T$.

\begin{figure*}[!t]
  \begin{minipage}[t]{0.5\linewidth}
    \centering
    \includegraphics[width=1.\linewidth]{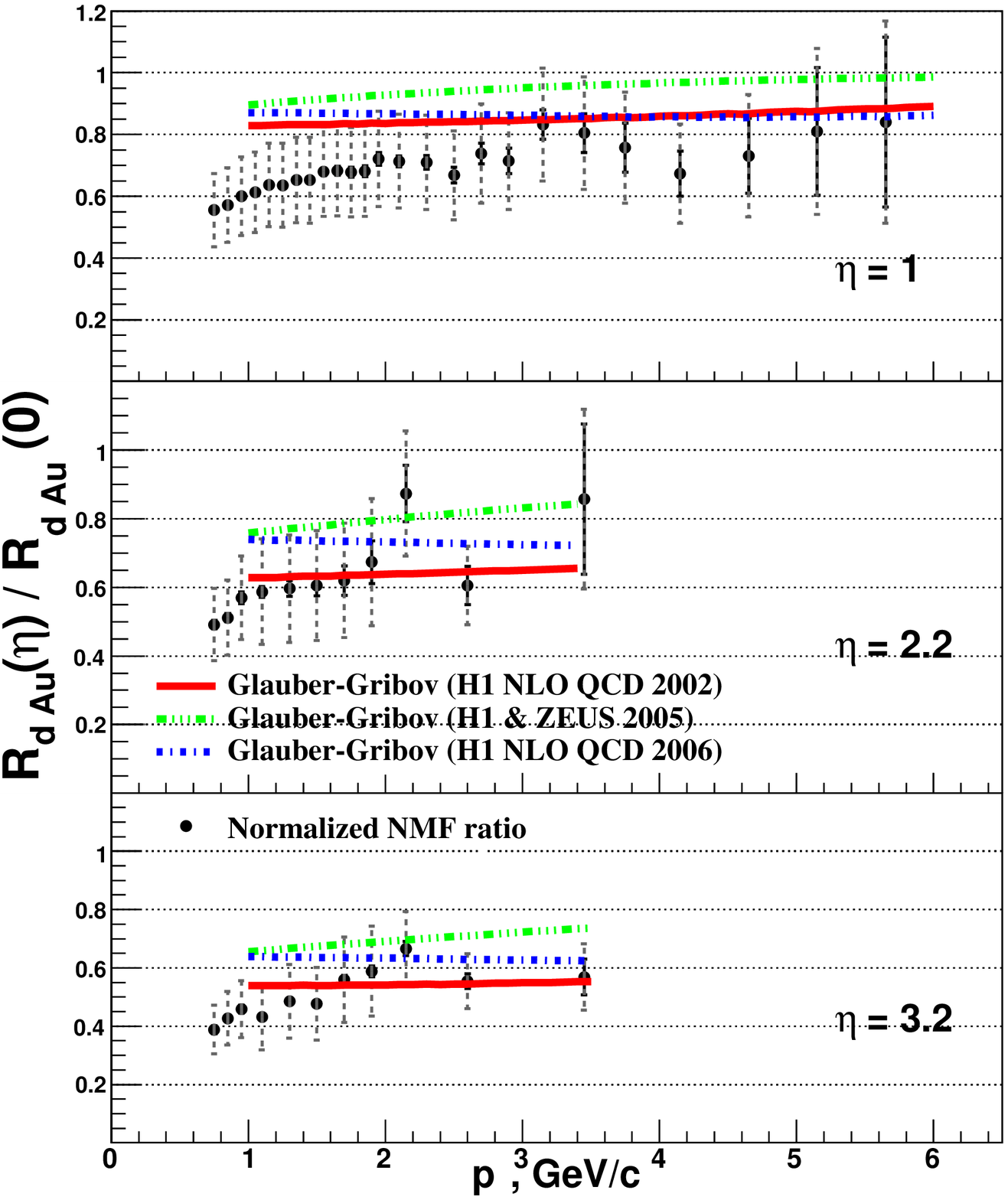}
    \caption{NMF ratio, $c = 3$}
    \label{fig:res_c3}
  \end{minipage}%
  \begin{minipage}[t]{0.5\linewidth}
    \centering
    \includegraphics[width=1.\linewidth]{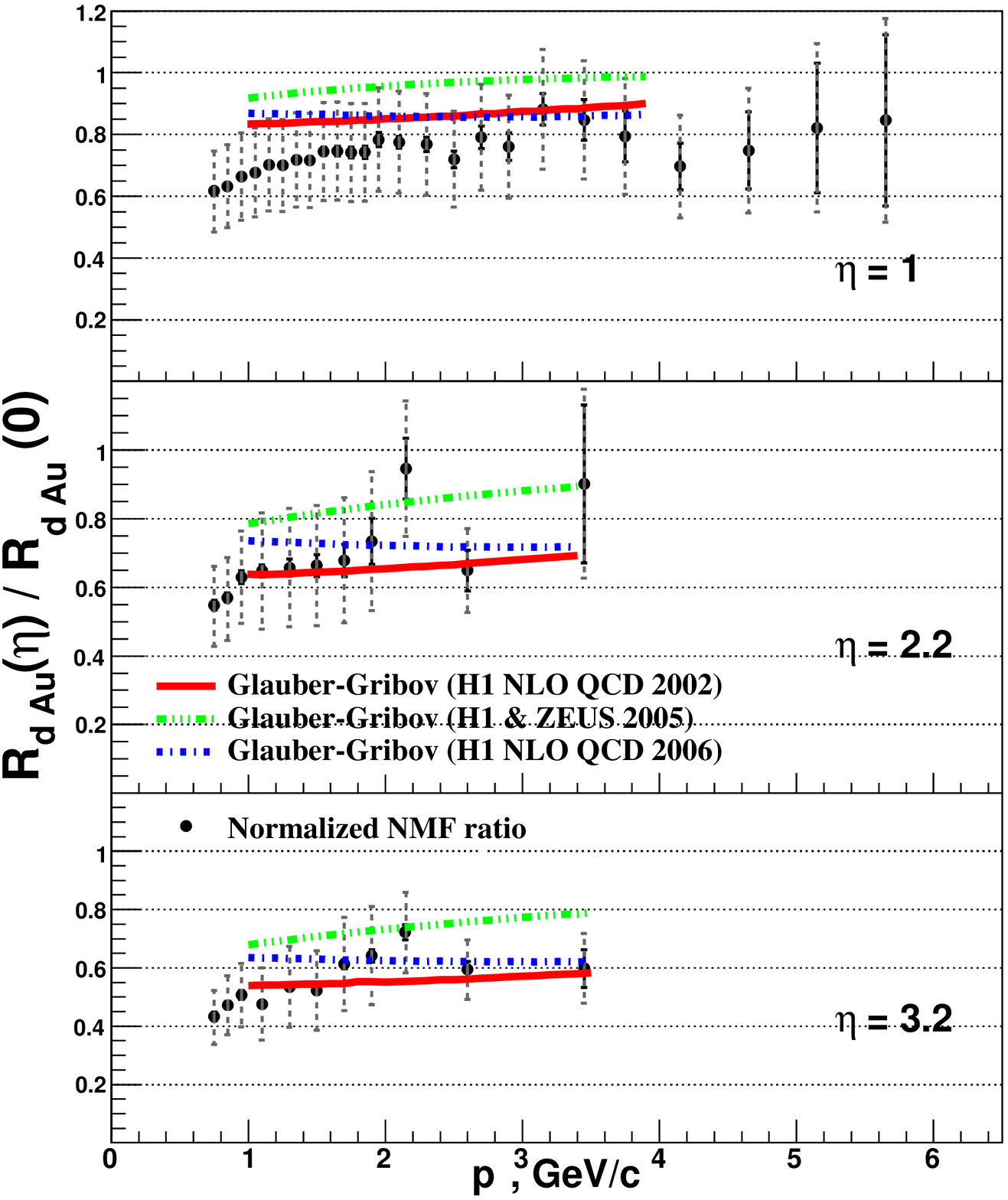}
    \caption{NMF ratio, $c = 5$}
    \label{fig:res_c5}
  \end{minipage}
\end{figure*}

The multiplicity reduction merely due to shadowing will then appear
as we compare the NMF at forward rapidities, $\eta = 1, \,2.2,
\,3.2$ to
 $R^{\mathit{norm}}_{d Au}$. This can be quantified by the double ratio
\beq
\label{eq:doubleratio}
\tilde{R} \;=\; \frac{\left[R_{d Au}\right]_{\eta}}{
  R^{\mathit{norm}}_{d Au}} \;,
\eeq where the NMF in the numerator is taken at a constant rapidity
slice.

The double ratio, as defined in (\ref{eq:doubleratio}), for the NMF
taken from \cite{brahms}
is plotted together with the predictions of the Glauber-Gribov model
from eq. (\ref{eq:shadow})
in figs. \ref{fig:res_c3} and \ref{fig:res_c5}
for two different values of the parameter $c$. Calculations are made
for the three gluon dPDF parameterizations described in
Section~\ref{sec:exp}. Statistical errors are
 denoted by the thick solid line, while the systematic and statistical
 errors added up quadratically are denoted by the dashed line.
The choice of $c$ does not seem to affect the result.

There is good agreement with experimental data for all the gluon
dPDF parameterizations. The H1 NLO QCD 2002 fit seems to be most
consistent with the data, and the agreement is better at higher
values of pseudorapidity. Although the ZEUS fit (H1 \& ZEUS 2005) is
almost a factor of two smaller for the whole range of possible $x$,
the agreement is surprisingly good. This gives evidence for a
dominating shadowing contribution to the suppression of the NMF at
forward rapidities at RHIC energies.

\section{Nuclear shadowing at SPS}
\label{sec:sps} 
\begin{figure}[!t]
  \centering
  \includegraphics[width=0.8\linewidth]{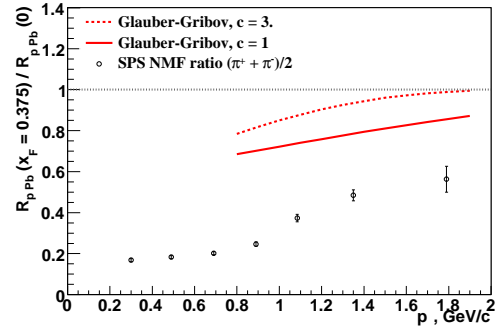}
  \caption{\label{fig:shadSPS} Shadowing ratio for $\sqrt{s} =
  17.3$GeV in the fragmentation region.}
\end{figure}

We have also calculated the nuclear shadowing ratio
for lower energies, namely at maximal SPS energy $\sqrt{s}=17.3$
GeV. In Fig.~\ref{fig:shadSPS} we show the calculation of the
Glauber-Gribov model compared to data on charged pion production for
fixed $x_F = 0.375$ taken from \cite{boimska} for the double ratio
defined in eq. (\ref{eq:doubleratio}) (the H1 NLO QCD 2002 fit was
used). The curves are for two values of the parameter $c$, and we
see that the shadowing disappears quickly with increasing $c$.

Obviously, the effect of gluon shadowing is not sufficient to
explain the observed suppression at this energy. The solid curve in
Fig.~\ref{fig:shadSPS} can be taken as the maximum value of the
effect. The suppression in the experimental data is strongest for
small $p_T$.

Theoretical considerations on the reasons for the suppression in the
NMF at SPS energies are out of the scope of the present paper, but
will be followed up in the nearest future \cite{oslo}. An extremely
interesting fact is that the suppression at SPS is almost of the
same magnitude as at RHIC energies. Since gluon shadowing is
expected to become significant with growing energy of the reaction,
there is apparently another mechanism present which is responsible
for the suppression. This mechanism is related to the energy-scale
relevant for coherent scattering; at these energies a large fraction
of the Fock-state of the incoming hadron will rescatter as in the
Glauber model \cite{kopeliovich}. Energy-momentum conservation
effects, which violate the AGK cutting rules, will play a dominant
role \cite{boreskovetal}.

\section{Conclusions}
\label{sec:concl} We have presented results for the nuclear
shadowing ratio, eq. (\ref{eq:sch}), and particle production in
hadron-nucleus collisions for various parameterizations of the gluon
dPDF within the Glauber-Gribov model, and compared them to
experimental data on the nuclear modification factor measured at two
different energies, $\sqrt{s} = 17.3$ GeV and $\sqrt{s} = 200$ GeV.
Our calculations of the nuclear shadowing ratio is consistent with
other models predicting large gluon shadowing for low values of $x$
and intermediate $Q^2$.

For RHIC energies, the suppression of the NMF observed at non-zero
values of pseudorapidity is well described by gluon shadowing within
the Glauber-Gribov model. The experimental data on gluon dPDF is
still quite uncertain, and introduces a large spread in the
theoretical prediction. The agreement with experimental data is
reasonable for all the presented parameterizations.

We would like to underline that the results we have presented can be
viewed as an upper bound of the effect of gluon shadowing in
hadron-nucleus collisions. The authors of \cite{gsv} have done
calculations within a similar framework with a different choice of
kinematics than in eq.~(\ref{eq:jet}), resulting in a much weaker
shadowing effect. This discrepancy is important to resolve in the
nearest future.

In \cite{gsv} there is also an important remark on the experimental
data from BRAHMS \cite{brahms} at the two most forward rapidities.
The fact that only negative particles, $h^-$, are measured leads
effectively to an enhancement of the NMF because of isospin effects.
In order to compare to the correct NMF, one should reduce the
experimental data at $\eta = 2.2$ and $3.2$ by a factor of $\sim
2/3$. This will lead to a less impressive agreement of our model,
yet it will not change the conclusions regarding gluon shadowing at
RHIC.

At SPS, the gluon shadowing is not responsible for more than 10 \%
of the total suppression. The important fact, however, is the
observed large suppression up to $p_T = 2$ GeV/c at such low
energies. The suppression seems in fact to be approximately of the
same magnitude as at RHIC, where $\sqrt{s}$ is a factor of ten
larger. The suppression is caused by energy-momentum conservation
which should stay constant with growing energy. This may indicate
that our estimates of gluon shadowing are too large.

The energy-dependence of the suppression is related to the
underlying space-time dynamics of the collision, and is therefore a
crucial test for theoretical models. With new low-energy data on NMF
in the forward region a comparison of the effect can now also be
done for $x_F > 0$. This gives an opportunity to study the interplay
of different effects leading to suppression/enhancement of particle
spectra in much more detail.

The presented model can also be used to calculate the expected
suppression in heavy-ion collisions at LHC energies \cite{oslo}.

\begin{acknowledgement}

K.~T. would like to thank for warm hospitality at SINP in Moscow
State University and ITEP, where part of this work has been done.
The authors would also like to thank B.~Boimska for providing
experimental data. Useful discussions with D.~R\"{o}hrich,
M.~Strikman and L.~Frankfurt are gratefully acknowledged. This work
has been supported by the agreement between the Department of
Physics, the University of Oslo and SINP MSU.

\end{acknowledgement}


\vfill\eject
\end{document}